\documentclass{INTERSPEECH2023}
\interspeechcameraready 

\usepackage{amsfonts}
\usepackage{url}
\usepackage{multicol}
\usepackage{multirow}
\usepackage{hyperref}
\usepackage{cleveref}
\usepackage[font=small,skip=0pt]{caption}
\usepackage{subfig}
\usepackage{xcolor} 
\usepackage{cite}
\usepackage{footmisc}

\title{Speech Intelligibility Assessment of Dysarthric Speech \\ by using Goodness of Pronunciation with Uncertainty Quantification}
\name{Eun Jung Yeo*$^1$\thanks{* Equal contribution.}, Kwanghee Choi*$^2$, Sunhee Kim$^1$, Minhwa Chung$^1$}
\address{
  $^1$Seoul National University, Republic of Korea\\
  $^2$Carnegie Mellon University, United States of America}
\email{ej.yeo@snu.ac.kr, kwanghec@andrew.cmu.edu, sunhkim@snu.ac.kr, mchung@snu.ac.kr}

\begin{document}

\maketitle

\begin{abstract}
This paper proposes an improved Goodness of Pronunciation (GoP) that utilizes Uncertainty Quantification (UQ) for automatic speech intelligibility assessment for dysarthric speech. Current GoP methods rely heavily on neural network-driven overconfident predictions, which is unsuitable for assessing dysarthric speech due to its significant acoustic differences from healthy speech. To alleviate the problem, UQ techniques were used on GoP by 1) normalizing the phoneme prediction (entropy, margin, maxlogit, logit-margin) and 2) modifying the scoring function (scaling, prior normalization). As a result, prior-normalized maxlogit GoP achieves the best performance, with a relative increase of 5.66\%, 3.91\%, and 23.65\% compared to the baseline GoP for English, Korean, and Tamil, respectively. Furthermore, phoneme analysis is conducted to identify which phoneme scores significantly correlate with intelligibility scores in each language.
\end{abstract}

\section{Introduction} \label{sec:intro}
Dysarthria is a motor speech disorder caused by weakness or paralysis of the articulators \cite{darley1972dysarthria}.
People with dysarthria often suffer from degraded speech intelligibility, repeated communication failures, and ultimately low quality of life. 
Accordingly, dysarthric speech assessments regarding speech intelligibility are conducted to check the patient's status and track the effectiveness of treatments \cite{kent1989toward}.
While the common way of dysarthric speech assessment is perceptual evaluation, the method is often subjective and laborious. 
Therefore, automatic speech assessment with objective and rapid results can assist clinicians in diagnosis and treatment planning.




There are two main approaches to automatic assessment
of dysarthric speech. 
The first approach is to propose a list of hand-crafted features that are expected to capture the characteristics of dysarthric speech.
Explored feature sets include voice quality features \cite{narendra2021automatic}, prosody features \cite{hernandez2020prosody}, articulation or pronunciation features \cite{kim2012automatic, liu2021language}, and their combinations \cite{vasquez2018towards, yeo2022multilingual}.
This approach has the benefit of having medical implications, as it provides a transparent understanding of the features employed for automatic assessment. 
Nevertheless, this approach has the drawback that features that could be valuable in the assessment could be discarded during feature extraction.
The second approach involves leveraging the capabilities of neural networks (NNs), which can achieve better results by using raw inputs. \cite{yeo2022automatic, joshy2023dysarthria}.
However, due to the black-box nature of NNs, the approach limits interpretability, which clinicians often crave for.

\begin{figure}[t] 
\centering
\includegraphics[width=0.97\columnwidth]{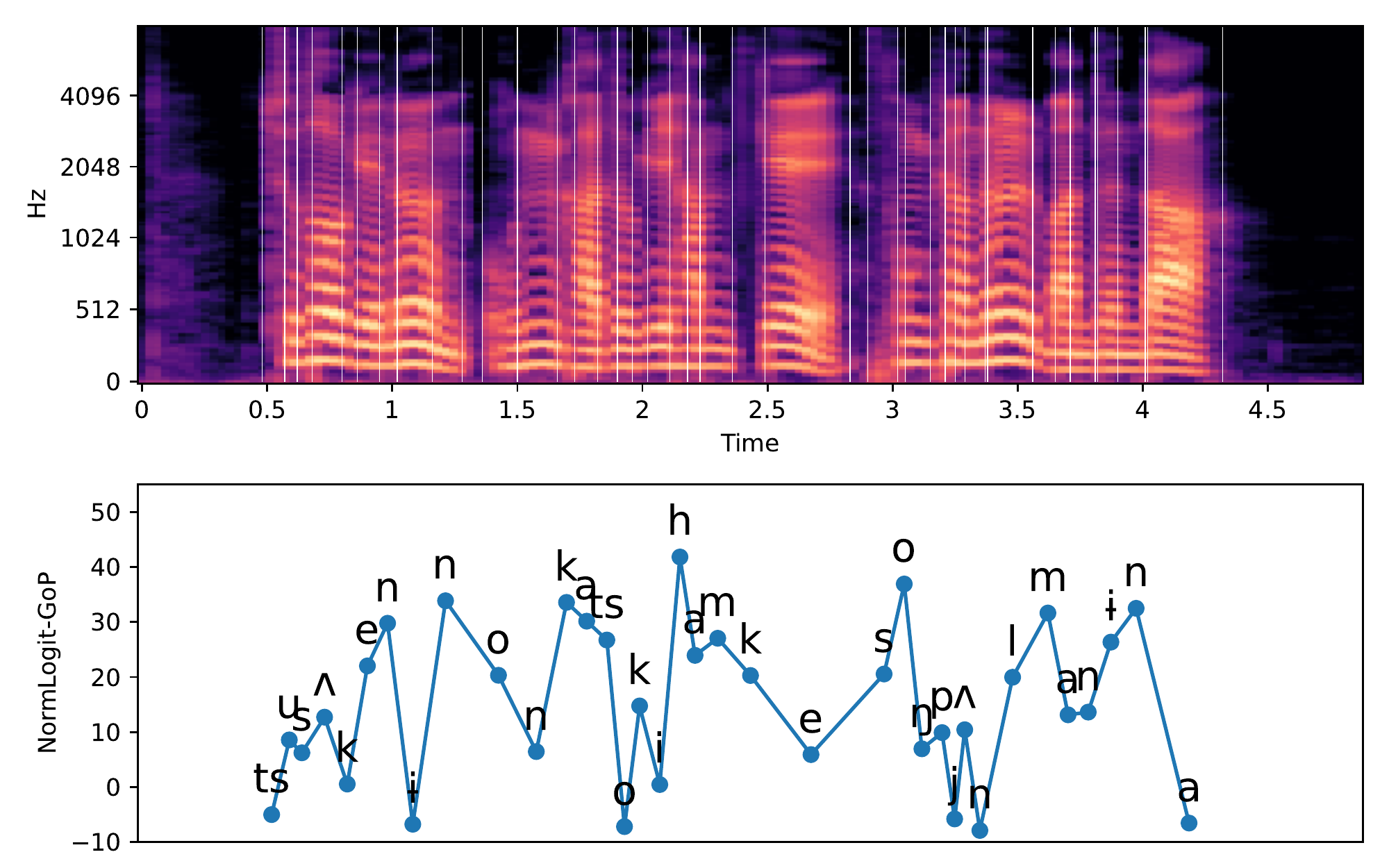}
\caption{
Example of GoP scores for each phone within an utterance.
Higher values indicate greater deviation from the correct pronunciation.
GoP scores allow easy identification of mispronounced phonemes.
}
\vspace{-1em}
\label{figs:samplewise_phones}
\end{figure}
Recent studies have attempted to integrate the benefits of both approaches, by enforcing the neural networks to learn the intermediate labels used for perceptual assessment, such as voice quality, articulation precision, nasality, and prosody \cite{tu2017interpretable, xu2021explore}.
Furthermore, certain studies focused on decoding misarticulation characteristics, which are the prominent aspect of dysarthric speech across languages \cite{yeo2022multilingual} and a significant factor that influences speech intelligibility \cite{de2002intelligibility}. 
For instance, the framework that measures the level of phonetic impairment was proposed by utilizing the activations of the hidden neuron \cite{abderrazek2022interpreting, abderrazek2022validation}.
While the method could provide overall phonetic characteristics of utterances, it is unable to provide assessments at the level of individual phonemes, which can help clinicians to pinpoint specific phonemes that require pronunciation training.



A common approach of phoneme-level speech assessment is to use the parallel NN that employs parallel datasets. 
The NN was trained using the same set of utterances recorded by both healthy speakers and patients to learn how to distinguish whether each phone in the utterances was from healthy or disordered speech \cite{miller2020assessing,quintas2022automatic}.
However, obtaining parallel datasets is a challenging task, especially for disordered speech.
Moreover, this approach often constrains the analysis to pre-defined speech materials, which may not capture the natural speech patterns utilized in everyday communication.


\begin{figure*}[t!] 
\centering
\includegraphics[width=1\textwidth,height=5cm]{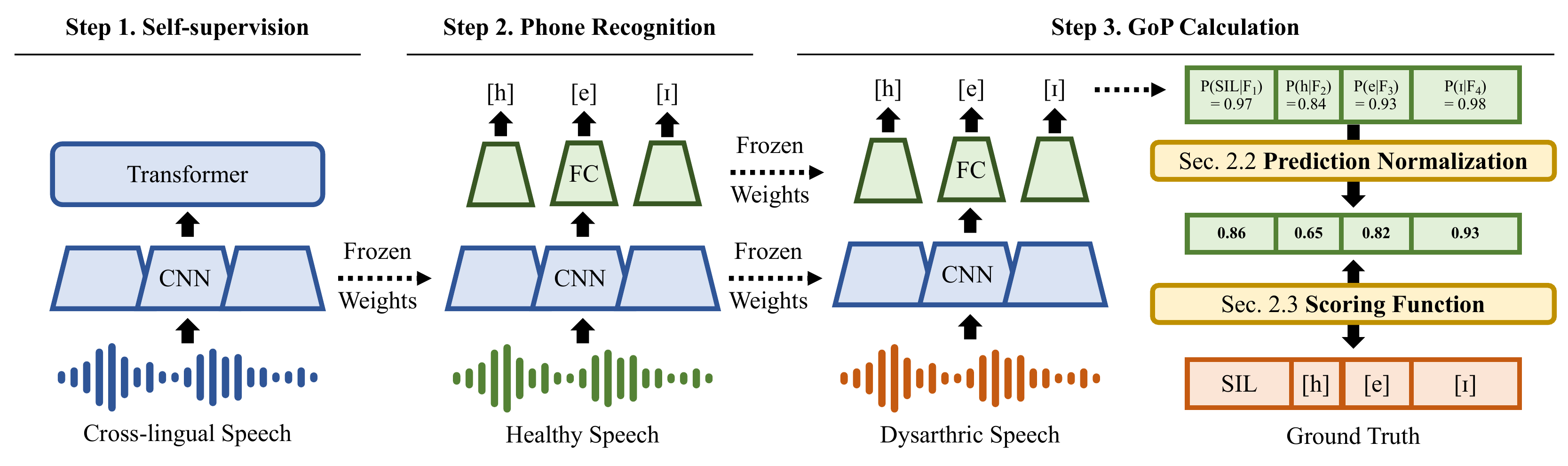}
\caption{Overview of the improved GoP with UQ methods.}
\vspace{-1em}
\label{fig:diagram}
\end{figure*}

Another conventional approach of phoneme-level pronunciation evaluation is Goodness of Pronunciation (GoP) \cite{witt2000phone}.
GoP, which is defined as the degree of similarity between produced and correct pronunciation of phonemes, has two advantages in automatic speech assessments.
First, it provides detailed information on which phonemes are mispronounced and to what extent each phoneme is atypical.
Second, it does not necessitate a parallel dataset for model training.
While GoP is often applied to non-native (L2) speech pronunciation assessment, some studies have also verified its potential use in assessing speech disorders as well \cite{pellegrini2014goodness,Fontan2015PredictingDS}.

With the development of NNs, variants of GoP which employ probabilities from the state-of-the-art neural networks have been suggested \cite{hu2015improved,Cheng2020ASRFreePA,xu2021explore}.
However, using these probabilities without taking into account the modern NNs' tendency towards overconfidence can result in inaccurate conclusions:
NNs often generate probabilities close to $1.0$ even when their predictions are incorrect \cite{guo2017calibration}.
Since GoP relies heavily on probabilities, this can be especially problematic. The issue is compounded when the model encounters out-of-distribution (OOD) inputs, which are data that differ significantly from the training data's distribution \cite{hendrycks2017baseline}, such as dysarthric speech for healthy speech.
To alleviate such issues, UQ techniques, techniques to combat OOD problem \cite{guo2017calibration,hendrycks2017baseline}, can be applied.

This paper proposes improved GoP for automatic speech intelligibility assessment for dysarthric speech by employing Uncertainty Quantification (UQ) methods in two ways: (1) normalizing the phoneme prediction and (2) modifying the scoring function.
As pathological speech greatly differs in acoustics from healthy speech \cite{wilson2000acoustic}, dysarthric speech can be also understood as OOD input.
Therefore, we employ conventional UQ methods to improve GoP calculations for dysarthric speech assessment.
To assess the effectiveness of the improved GoP with UQ techniques, three dysarthric datasets, namely, UASpeech English, QoLT Korean, and SSNCE Tamil dataset, are utilized. 
To summarize, this paper redefines the current versions of GoP from a UQ standpoint and evaluates the effectiveness of UQ methods in improving GoPs.

\section{Proposed approach} \label{sec:gop}
The study applies various conventional UQ methods to calculate GoP scores, which are demonstrated in \Cref{fig:diagram}.
We release the source code of all the experiments.\footnote{\url{https://github.com/juice500ml/dysarthria-gop}\label{footnote:repo_url}}

\subsection{Prerequisite: Goodness of Pronunciation (GoP)} \label{ssec:gop-baseline}
In this subsection, we provide a succinct summary of the existing GoP-based methods from the uncertainty quantification perspective.
Starting from GMM-GoP, \Cref{eq:gmm-gop} presents the definition of the corresponding method:
Given a phone $p$ during frames $f \in F$ with frame-wise phone probability and its logits as $P^f(p|f)$ and $L^f(p|f)$, and the total phone set as $Q$, \textbf{GMM-GoP} \cite{witt2000phone} is defined as an averaged log probability across the phone duration, $\mathbb{E}_F[\log P^f(p|f)]$:
\begin{equation}
    s_\text{GMM-GoP}(p) = \frac{1}{|F|} \sum_{f \in F} \log \frac{e^{L^f(p|f)}}{ \sum_{q \in Q} e^{L^f(q|f)}}.
\label{eq:gmm-gop}
\end{equation}
Averaging the log probabilities can be seen as a form of temporal ensembling \cite{laine2017temporal}, which is a well-known UQ method, as it combines the predictions from multiple frames into a single estimate. 
Further, directly using the probability output is often used as a baseline for OOD \cite{hendrycks2017baseline}.

Another popular baseline is \textbf{NN-GoP} \cite{hu2015improved}, which improved GMM-GoP by leveraging the development of deep neural networks and modifying the score function:
\begin{align}
    \Bar{P}(p|F) = \mathbb{E}_F[P^f(p|f)] = \frac{1}{|F|} \sum_{f \in F} P^f(p|f), \\ 
    s_\text{NN-GoP}(p) = \log \Bar{P}(p|F) - \max_{q \in Q} \log \Bar{P}(q|F). \label{eq:nn-gop}
\end{align}

Finally, \textbf{DNN-GoP} \cite{hu2015improved} normalizes the phone probability with the phone prior \cite{hong2021disentangling}:
\begin{align}
    s_\text{DNN-GoP}(p) = \Bar{P}(p|F) / P(p).
\end{align}




\subsection{Normalizing the phoneme predictions}\label{subsec:modify_pred}
There are two commonly used ways to calibrate the posterior prediction $P(p|F)$ by modifying its logit $L(p|F)$.
One is to normalize by removing the influence of the prior $P(p)$ \cite{hong2021disentangling,hu2015improved} (\textbf{Prior}), and the other is to reduce the peakiness by temperature scaling \cite{guo2017calibration} \textbf{Scale}:
\begin{align}
    L^f(p, f) &= L^f(p|f) - \log P(p), \\
    L^f_T(p|f) &= L^f(p|f) / T, 
\end{align}
where $T$ is the hyperparameter and the modified predictions are the softmax function's output of the corresponding logits.

Normalizing via the prior is the same with the idea of DNN-GoP, where it is commonly applied to disentangle the training distribution of the phone recognizer, where majority classes are often overconfident than minority classes \cite{hong2021disentangling}.
Temperature scaling is also commonly used as a baseline for UQ, as it avoids the peaky distribution of posterior probabilities.

\subsection{Modifying the scoring function}\label{subsec:modify_score}
We first employ one of the most common methods to measure the data uncertainty: Entropy $H$ and Margin $M$.
\textbf{Entropy} measures the uncertainty associated with a set of possible outcomes:
\begin{align}
    s_\text{H}(p) = -\sum_{q \in Q} \Bar{P}(q|F) \log \Bar{P}(p|F).
\label{eq:entropy-gop}
\end{align} 
Specifically, the entropy represents the average amount of information needed to specify which outcome was actually observed.
Entropy does not require ground truth labels, so it is often used when obtaining the labels is expensive.

\textbf{Margin} refers to the difference between the true class and the highest class probabilities for a given prediction:
\begin{align}
    s_\text{M}(p) = \Bar{P}(p|F) - \max_{q \in Q - \{p\}} \Bar{P}(q|F).
\label{eq:margin-gop}
\end{align}
Note that the equation is strikingly similar to that of NN-GoP except the above definition excludes the true phone as $Q - \{p\}$.

On the other hand, the \textbf{MaxLogit} \cite{hendrycks2022scaling} and \textbf{LogitMargin} involves directly utilizing the logits.
Softmax function is often known to squash the useful information inside logits, so that it can normalize the sum into one.
Hence, one can apply the idea of both GMM-GoP (directly using the probability) and NN-GoP (using the Margin):
\begin{align}
    s_\text{MaxLogit}(p) &= \frac{1}{|F|} \sum_{f \in F} L^f(p|f) = \Bar{L}(p|F), \\
    s_\text{LogitMargin}(p) &= \Bar{L}(p|F) - \max_{q \in Q - \{p\}} \Bar{L}(p|F).
\label{eq:logit-gop}
\end{align}

\section{Experimental setting} \label{sec:experiment}

\subsection{Datasets} \label{ssec:corpus}
To train the acoustic model to learn the distribution of healthy phonemes, we use the Common Phone dataset \cite{klumpp2022common} and the L2-ARCTIC dataset \cite{zhao2018l2}.
To evaluate the efficacy of our proposed approach, three dysarthric datasets are utilized: UASpeech English dataset \cite{kim2008uaspeech}, QoLT Korean dataset \cite{choi2012dysarthric}, and SSNCE Tamil dataset \cite{ta2016dysarthric}.
All three datasets contain dysarthric speech from speakers suffering from Cerebral Palsy.
We focus the analysis on sentences from QoLT and SSNCE dysarthric datasets, since the Common Phone and L2-ARCTIC datasets solely consist of sentences.
Word materials are analyzed for the UASpeech dataset, since the dataset contains words only.

\subsubsection{Common Phone dataset and L2-ARCTIC dataset} \label{sssec:commonPhone}
The acoustic model is trained on healthy speech using the Common Phone dataset and the L2-ARCTIC dataset, which were selected for their extensive phoneme coverage and phonetic annotations.
These datasets are expected to cover most of the phonemes used in English, Korean, and Tamil.
\textbf{Common Phone dataset} \cite{klumpp2022common} is a gender-balanced, multilingual corpus with six languages.
Comprised of more than 11,000 speakers, the dataset includes around 116 hours of speech.
\textbf{L2-ARCTIC dataset} \cite{zhao2018l2} is a speech corpus often used for detecting mispronunciations in non-native English speakers. 
It includes recordings from 24 speakers with a balanced distribution of gender and first language, representing six different countries. 
On average, each speaker has around 67.7 minutes of speech, which has a total duration of approximately 27.1 hours.

\subsubsection{UASpeech English dysarthric datasat} \label{sssec:uaspeech}
UASpeech dataset \cite{kim2008uaspeech} is a publicly-available English dysarthric speech dataset, which contains 15 dysarthria speakers (11 males, 4 females) and 13 aged-matched healthy speakers (9 males, 4 females). 
Speakers were classified based on the scores on the Frenchay Dysarthria Assessment (FDA) \cite{enderby1980frenchay}: 5 mild speakers (score 1), 3 moderate-to-severe speakers (score 2), 3 moderate-to-severe speakers (score 3), and 4 severe speakers (score 4). 
Montreal Forced Aligner (MFA) \cite{mcauliffe2017montreal} is employed to extract phoneme-level alignments.



\subsubsection{QoLT Korean dysarthric dataset} \label{sssec:qolt}
Quality of Life Technology (QoLT) dataset \cite{choi2012dysarthric} is a privately held dataset of Korean dysarthric speech. 
The corpus consists of 70 dysarthric speakers (45 males, 25 females) and 10 healthy speakers (5 males, 5 females).
Each speaker recorded five phonetically balanced sentences twice. 
Five speech pathologists were asked to determine the intelligibility levels of the speakers on a 5-point Likert scale.
With a score of 0 considered healthy, the dataset holds 25 mild (score 1), 26 mild-to-moderate (score 2), 12 moderate-to-severe (score 3), and 7 severe (score 4) intelligibility level speakers.
Accordingly, 100 healthy utterances and 700 dysarthric utterances are used for the experiment.
After using MFA to align the phonemes, two speech pathologists further fixed the automated alignment for better quality.

\subsubsection{SSNCE Tamil dysarthric dataset} \label{sssec:ssnce}
SSNCE dataset \cite{ta2016dysarthric} is a Tamil dysarthric speech corpus available by request. 
The dataset includes recordings from 20 dysarthric speakers (13 males and 7 females) and 10 healthy speakers (5 males and 5 females). 
The dataset groups dysarthric speakers based on their speech intelligibility scores, which were marked by two speech pathologists on a 7-point Likert scale.
A score of 0 considered healthy, score 1 and 2 are grouped into mild (score 1), score 3 and 4 into moderate (score 2), and score 5 and 6 into severe (score 3).
There were different numbers of speakers in each score category: 7 with mild, 10 with moderate, and 3 with severe.
For the experiment, we used a total of 5,200 utterances from the dysarthric speakers and 2,600 utterances from the healthy speakers, with 260 unique sentences recorded from each speaker.
For forced alignments, we use the time-aligned phonetic transcriptions provided by the dataset.

\subsection{Experimental details} \label{ssec:detail}
In this study, we evaluated GoP performance by following the approach of the previous study \cite{Fontan2015PredictingDS}.
Concretely, rather than evaluating the models based on their accuracy in mispronunciation detection, our objective was to calculate the average GoP score for each utterance and compare their correlation with the intelligibility scores.

\subsubsection{Phoneme Prediction} \label{sssec:phone_pred}
To fairly compare the performances between various GoP scoring functions, we extract the posterior probabilities from the common cross-lingual Wav2Vec 2.0 XLS-R model \cite{babu2022xls} instead of using the acoustic model from Kaldi, following recent literature \cite{xu2021explore}.
We slightly modify the architecture by attaching the linear phone prediction head to the convolutional layer, not the transformer layer, to avoid extensive computational overhead and preserve the phonetic characteristics in convolutional features \cite{choi2022opening}.
Also, the loss function is simplified by removing the adaptive pooling, where we observed that the final performance difference was negligible.
AdamW optimizer \cite{loshchilov2019decoupled} is used with the default learning rate of $0.001$ for three epochs.
As we only trained the linear prediction head, the final performance was not sensitive to other hyperparameters.
Refer to \cite{xu2021explore} and our source code\footref{footnote:repo_url} for more details.


\subsubsection{Baselines} \label{sssec:baselines}
We conducted three baseline experiments: GMM-GoP \cite{witt2000phone}, NN-GoP \cite{hu2015improved}, and DNN-GoP.
We compare the baselines with the UQ methods introduced in \Cref{subsec:modify_score} and \Cref{subsec:modify_pred} by using the same phoneme probabilities.
As we aim to see the correlations between GoP scores (continuous) and intelligibility scores (ordinal),
we utilize the Kendall Rank Coefficient $\tau$ to compare the performances.
Kendall's $\tau$ measures the strength of the relationships, with a higher absolute coefficient indicating higher correlations between the two variables.

\begin{table}[t]
\caption{
Kendall's rank coefficient between GoP \& intelligibility severity levels. A higher absolute value indicates a stronger correlation between the two variables.
}
\vspace{1em}
\label{tab:results}
\centering
\resizebox{0.48\textwidth}{!}
{

\begin{tabular}{c|c|c|c|c|c}
\hline
Method & Norm. & Scoring Func. & English & Korean & Tamil \\
\hline
\multirow{3}{*}{Baseline} 
& None & GMM \cite{witt2000phone,hmm-gop} & -0.2049 & -0.5237 & -0.3571 \\
& None & NN \cite{hu2015improved} & -0.1536 & -0.4687 & -0.4003 \\
& Prior & DNN-GoP \cite{hu2015improved} & -0.1836 & -0.4237 & -0.4681 \\
\hline
\multirow{12}{*}{Proposed} 
& \multirow{4}{*}{None}  & Entropy & -0.1831 & -0.2643 & -0.3251  \\
&  & Margin & -0.1628 & -0.4434 & -0.4445 \\
&  & MaxLogit & -0.2164 & -0.5440 & -0.5786 \\
&  & LogitMargin & -0.1732 & -0.4753 & -0.5158 \\
\cline{2-6}
& \multirow{4}{*}{Scale} & Entropy & -0.1755 & -0.1974 & -0.2263 \\
&  & Margin & -0.1260 & -0.4444 & -0.4210\\
&  & MaxLogit & -0.2164 & -0.5440 & -0.5786 \\
&  & LogitMargin & -0.1732 & -0.4753 & -0.5158\\
\cline{2-6}
& \multirow{4}{*}{Prior} & Entropy & -0.1833 & -0.2645 & -0.3254  \\
&  & Margin & -0.1630 & -0.4432 & -0.4447\\
&  & \textbf{MaxLogit} & \textbf{-0.2165} & \textbf{-0.5442} & \textbf{-0.5788}\\
&  & LogitMargin & -0.1733 & -0.4753 & -0.5160\\
\hline
\end{tabular}
}
\vspace{-1em}
\end{table}

%
\section{Experimental results} \label{sec:results}
\subsection{Correlation between GoPs and intelligilbity scores}
\Cref{tab:results} demonstrates the performances of both baseline and proposed experiments, with the best performance indicated in bold.
GoP with prior normalized MaxLogit performed the best on all the languages among the baselines and the UQ methods, achieving $-21.65\%$, $-54.42\%$, $-57.88\%$ correlation, for English, Korean, and Tamil, respectively.

The results of the baseline experiments show that GMM-GoP has the highest correlation for English and Korean at $-20.49\%$ and $-52.37\%$, respectively, while DNN-GoP performs best for Tamil with a correlation of $-46.81\%$.
For the proposed experiments, GoP without normalization generally shows lower performance than the baseline, except for MaxLogit-based GoP. 
Additionally, while scaling normalization has minimal impact, prior normalization has a positive effect on GoP performance for all languages. 
Furthermore, when entropy-based, probability-based (Margin), and logit-based (MaxLogit, LogitMargin) GoP variants are compared, the logit-based GoPs show the highest correlations to the intelligibility scores.
Additionally, performances on English are notably lower than that of Tamil and Korean.
We suspect that automatically generated alignment causes the degradation to occur \cite{mathad2021impact}, for the severe cases where alignment becomes much more challenging.
We aim to mitigate this issue in our future work.

\subsection{Analysis on phonemes}
\begin{figure}[t] 
\centering
\subfloat[Korean phoneme \textipa{/i/}.]{\includegraphics[width=0.43\columnwidth]{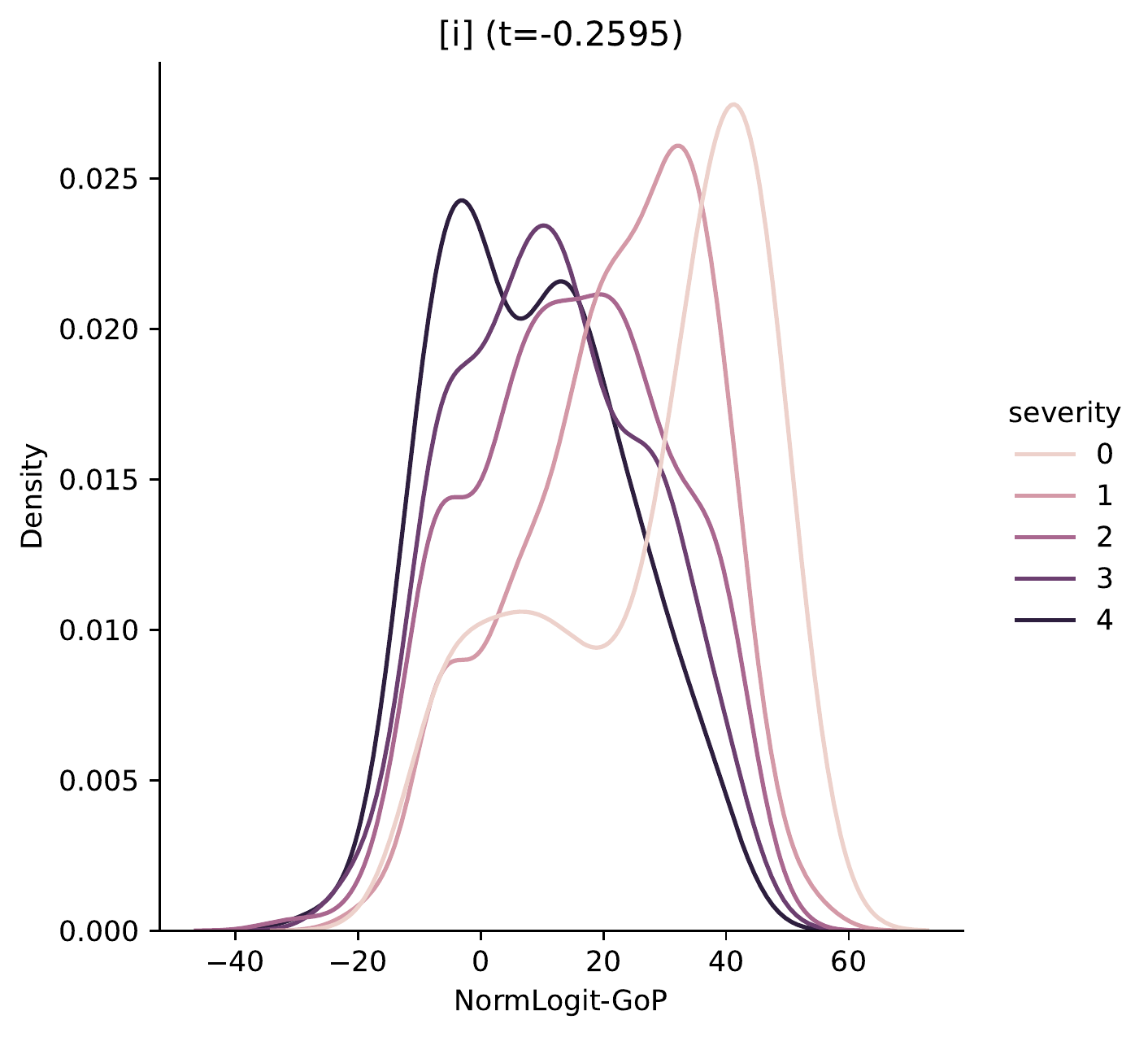}}
\subfloat[Korean phoneme \textipa{/m/}.]{\includegraphics[width=0.43\columnwidth]{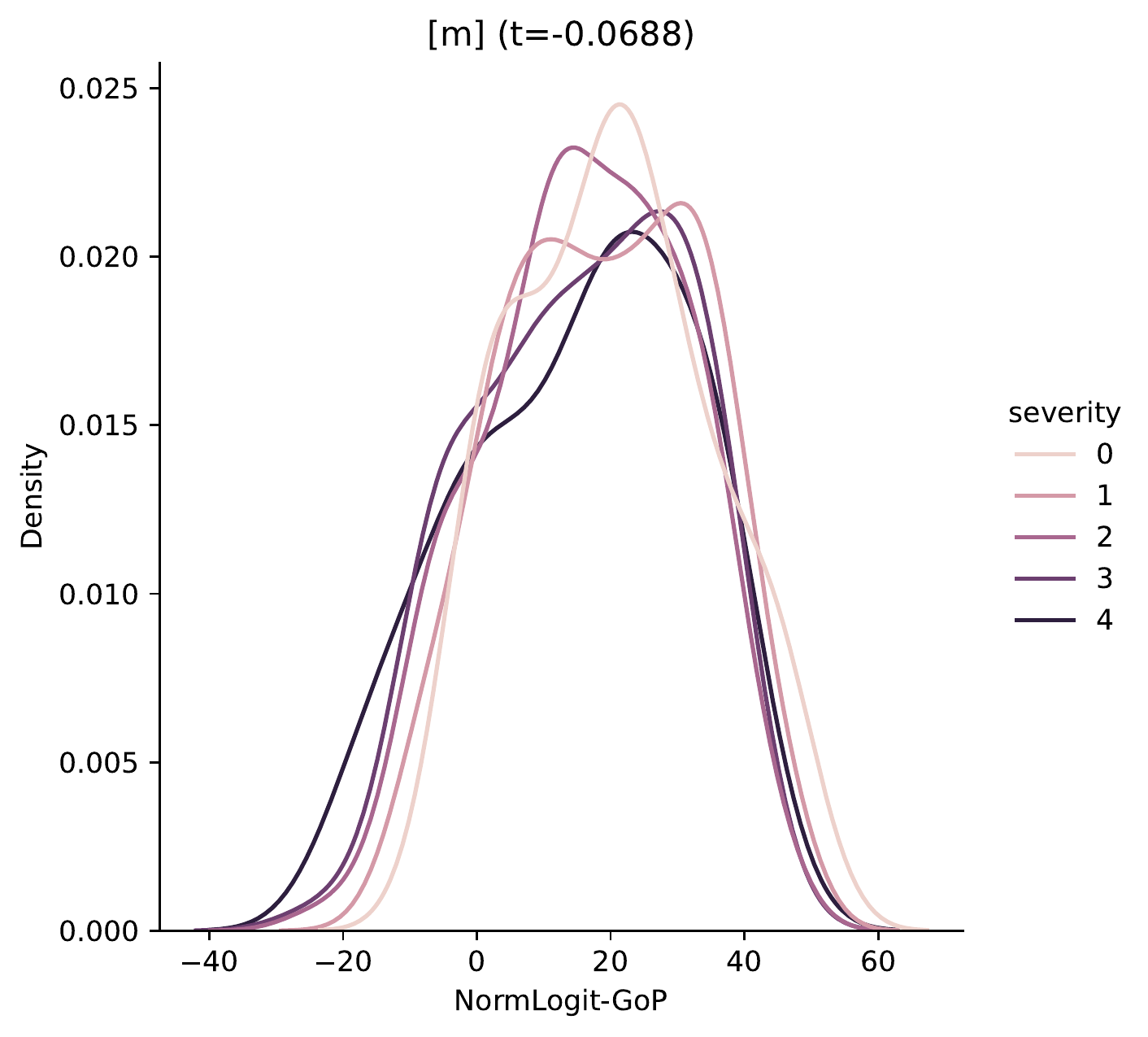}}
\vspace{1em}
\caption{
Kendall's $\tau$ distributions for two phonemes by severity.
0:healthy, 1:mild, 2:mild-to-mod., 3:mod.-to-sev., 4:severe.
}
\vspace{-1em}
\label{figs:phones}
\end{figure}

\Cref{figs:phones} illustrates the GoP distribution between two Korean phonemes \textipa{/i/} and \textipa{/m/}.
While the distribution of \textipa{/i/} differs significantly, the distribution of \textipa{/m/} is similar across all severity levels.
This finding suggests that certain phonemes have more impact power for severity levels based on speech intelligibility, which is consistent with previous findings \cite{quintas2022automatic}.

Identifying which phoneme pronunciation scores highly correlate to speech intelligibility can be useful in the clinical scenario, such as diagnosis and treatment.
For example, as demonstrated in \Cref{figs:samplewise_phones}, one can pinpoint the mispronounced phonemes within a single utterance.

Further, we conduct a quantitative analysis on which phoneme scores highly correlate to speech intelligibility by languages.
Kendall's $\tau$ is calculated for each phoneme between our best-performing Prior+MaxLogit GoP score and the intelligibility scores.
The top-5 phonemes based on correlation are as follows- English: \textipa{/a\textsci/},\textipa{/\textesh/},\textipa{/a\textupsilon/},\textipa{/z/},\textipa{/\textdyoghlig/}; Korean:\textipa{/i/},\textipa{/s/},\textipa{/n/},\textipa{/a/},\textipa{/\textturnv/}; Tamil:\textipa{/\textrtails/},\textipa{/h/},\textipa{/\textteshlig/},\textipa{/z/},\textipa{/a\textsci/}.
In summary, fricative sounds (\textipa{/s/},\textipa{/\textrtails/},\textipa{/\textesh/},\textipa{/z/}) strongly correlates to speech intelligibility across languages, consistent with the previous results \cite{hernandez2019acoustic}.
Affricates (\textipa{/\textteshlig/},\textipa{/\textdyoghlig/}) and diphthongs (\textipa{/a\textsci/},\textipa{/a\textupsilon/}) were shared as the top-5 phoneme list for English and Tamil.
This can be explained by the complexity in the articulation of affricates and diphthongs leading to difficulties in correct pronunciation for speakers with lower speech intelligibility.
On the other hand, Korean showed higher correlations for nasal (\textipa{/n/}) and monophthongs (\textipa{/a/},\textipa{/i/},\textipa{/\textturnv/}).
This may be again related to the movement of the articulators, such as the tongue, velum, and jaw.
We additionally provide the correlation scores of all the phonemes in our repository.\footref{footnote:repo_url}

\section{Conclusion} \label{sec:conclusion}
This paper proposes an improved GoP for dysarthria speech intelligibility assessment by using UQ methods.
Expected to alleviate the problem of modern NN's overconfidence, especially for disordered speech, tested UQ methods include (1) normalization of phoneme prediction and (2) modification of the scoring function. 
The experiments were carried out on dysarthric speech datasets in English, Korean, and Tamil.
According to the experimental results, the prior normalized MaxLogit GoP shows the best performance, outperforming both the traditional GoPs and other proposed GoP variants.
Furthermore, to verify the usefulness of our proposed method, an analysis of which phoneme pronunciation scores highly correlate to speech intelligibility is conducted.

\section{Acknowledgements}
This work was supported by Institute of Information \& communications Technology Planning \& Evaluation (IITP) grant funded by the Korea government (MSIT) (No.2022-0-00223, Development of digital therapeutics to improve communication ability of autism spectrum disorder patients).

\bibliographystyle{IEEEtran}
\bibliography{mybib}

\begin{thebibliography}{10}
\providecommand{\url}[1]{#1}
\csname url@samestyle\endcsname
\providecommand{\newblock}{\relax}
\providecommand{\bibinfo}[2]{#2}
\providecommand{\BIBentrySTDinterwordspacing}{\spaceskip=0pt\relax}
\providecommand{\BIBentryALTinterwordstretchfactor}{4}
\providecommand{\BIBentryALTinterwordspacing}{\spaceskip=\fontdimen2\font plus
\BIBentryALTinterwordstretchfactor\fontdimen3\font minus
  \fontdimen4\font\relax}
\providecommand{\BIBforeignlanguage}[2]{{%
\expandafter\ifx\csname l@#1\endcsname\relax
\typeout{** WARNING: IEEEtran.bst: No hyphenation pattern has been}%
\typeout{** loaded for the language `#1'. Using the pattern for}%
\typeout{** the default language instead.}%
\else
\language=\csname l@#1\endcsname
\fi
#2}}
\providecommand{\BIBdecl}{\relax}
\BIBdecl

\bibitem{darley1972dysarthria}
F.~L. Darley, J.~R. Brown, and N.~P. Goldstein, ``Dysarthria in multiple
  sclerosis,'' \emph{Journal of Speech and Hearing research}, vol.~15, no.~2,
  pp. 229--245, 1972.

\bibitem{kent1989toward}
R.~D. Kent, G.~Weismer, J.~F. Kent, and J.~C. Rosenbek, ``Toward phonetic
  intelligibility testing in dysarthria,'' \emph{Journal of Speech and Hearing
  Disorders}, vol.~54, no.~4, pp. 482--499, 1989.

\bibitem{narendra2021automatic}
N.~Narendra and P.~Alku, ``Automatic assessment of intelligibility in speakers
  with dysarthria from coded telephone speech using glottal features,''
  \emph{Computer Speech \& Language}, vol.~65, p. 101117, 2021.

\bibitem{hernandez2020prosody}
A.~Hernandez, S.~Kim, and M.~Chung, ``Prosody-based measures for automatic
  severity assessment of dysarthric speech,'' \emph{Applied Sciences}, vol.~10,
  no.~19, p. 6999, 2020.

\bibitem{kim2012automatic}
M.~J. Kim and H.~Kim, ``Automatic assessment of dysarthric speech
  intelligibility based on selected phonetic quality features,'' in
  \emph{ICCHP}.\hskip 1em plus 0.5em minus 0.4em\relax Springer, 2012, pp.
  447--450.

\bibitem{liu2021language}
Y.~Liu, N.~Penttil{\"a}, T.~Ihalainen, J.~Lintula, R.~Convey, and
  O.~R{\"a}s{\"a}nen, ``Language-independent approach for automatic computation
  of vowel articulation features in dysarthric speech assessment,''
  \emph{TASLP}, vol.~29, pp. 2228--2243, 2021.

\bibitem{vasquez2018towards}
J.~C. V{\'a}squez-Correa, J.~Orozco-Arroyave, T.~Bocklet, and E.~N{\"o}th,
  ``Towards an automatic evaluation of the dysarthria level of patients with
  parkinson's disease,'' \emph{Journal of communication disorders}, vol.~76,
  pp. 21--36, 2018.

\bibitem{yeo2022multilingual}
E.~J. Yeo, S.~Kim, and M.~Chung, ``Multilingual analysis of intelligibility
  classification using english, korean, and tamil dysarthric speech datasets,''
  in \emph{Oriental-COCOSDA}, 2022, pp. 1--6.

\bibitem{yeo2022automatic}
E.~J. Yeo, K.~Choi, S.~Kim, and M.~Chung, ``Automatic severity assessment of
  dysarthric speech by using self-supervised model with multi-task learning,''
  in \emph{ICASSP}, 2023.

\bibitem{joshy2023dysarthria}
A.~A. Joshy and R.~Rajan, ``Dysarthria severity classification using multi-head
  attention and multi-task learning,'' \emph{Speech Communication}, vol. 147,
  pp. 1--11, 2023.

\bibitem{tu2017interpretable}
M.~Tu, V.~Berisha, and J.~Liss, ``Interpretable objective assessment of
  dysarthric speech based on deep neural networks.'' in \emph{Interspeech},
  2017, pp. 1849--1853.

\bibitem{xu2021explore}
X.~Xu, Y.~Kang, S.~Cao, B.~Lin, and L.~Ma, ``Explore wav2vec 2.0 for
  mispronunciation detection.'' in \emph{Interspeech}, 2021, pp. 4428--4432.

\bibitem{de2002intelligibility}
M.~S. De~Bodt, M.~E. H.-D. Huici, and P.~H. Van De~Heyning, ``Intelligibility
  as a linear combination of dimensions in dysarthric speech,'' \emph{Journal
  of communication disorders}, vol.~35, no.~3, pp. 283--292, 2002.

\bibitem{abderrazek2022interpreting}
S.~Abderrazek, C.~Fredouille, A.~Ghio, M.~Lalain, C.~Meunier, and V.~Woisard,
  ``Interpreting deep representations of phonetic features via neuro-based
  concept detector: Application to speech disorders due to head and neck
  cancer,'' \emph{TASLP}, vol.~31, pp. 200--214, 2022.

\bibitem{abderrazek2022validation}
------, ``Validation of the neuro-concept detector framework for the
  characterization of speech disorders: A comparative study including
  dysarthria and dysphonia,'' in \emph{Interspeech}, 2022.

\bibitem{miller2020assessing}
G.~F. Miller, J.~C. V{\'a}squez-Correa, and E.~N{\"o}th, ``Assessing the
  dysarthria level of parkinson’s disease patients with gmm-ubm supervectors
  using phonological posteriors and diadochokinetic exercises,'' in \emph{TSD},
  2020, pp. 356--365.

\bibitem{quintas2022automatic}
S.~Quintas, J.~Mauclair, V.~Woisard, and J.~Pinquier, ``Automatic assessment of
  speech intelligibility using consonant similarity for head and neck cancer,''
  in \emph{Interspeech}, 2022.

\bibitem{witt2000phone}
S.~M. Witt and S.~J. Young, ``Phone-level pronunciation scoring and assessment
  for interactive language learning,'' \emph{Speech communication}, vol.~30,
  no. 2-3, pp. 95--108, 2000.

\bibitem{pellegrini2014goodness}
T.~Pellegrini, L.~Fontan, J.~Mauclair, J.~Farinas, and M.~Robert, ``The
  goodness of pronunciation algorithm applied to disordered speech,'' in
  \emph{Fifteenth Annual Conference of the International Speech Communication
  Association}, 2014.

\bibitem{Fontan2015PredictingDS}
L.~Fontan, T.~Pellegrini, J.~Olcoz, and A.~Abad, ``Predicting disordered speech
  comprehensibility from goodness of pronunciation scores,'' in
  \emph{SLPAT@Interspeech}, 2015.

\bibitem{hu2015improved}
W.~Hu, Y.~Qian, F.~K. Soong, and Y.~Wang, ``Improved mispronunciation detection
  with deep neural network trained acoustic models and transfer learning based
  logistic regression classifiers,'' \emph{Speech Communication}, vol.~67, pp.
  154--166, 2015.

\bibitem{Cheng2020ASRFreePA}
S.~Cheng, Z.~Liu, L.~Li, Z.~Tang, D.~Wang, and T.~F. Zheng, ``Asr-free
  pronunciation assessment,'' in \emph{Interspeech}, 2020.

\bibitem{guo2017calibration}
C.~Guo, G.~Pleiss, Y.~Sun, and K.~Q. Weinberger, ``On calibration of modern
  neural networks,'' in \emph{ICML}, 2017, pp. 1321--1330.

\bibitem{hendrycks2017baseline}
D.~Hendrycks and K.~Gimpel, ``A baseline for detecting misclassified and
  out-of-distribution examples in neural networks,'' in \emph{ICLR}, 2017.

\bibitem{wilson2000acoustic}
J.~Wilson, Bronagh~Blaney, ``Acoustic variability in dysarthria and computer
  speech recognition,'' \emph{Clinical Linguistics \& Phonetics}, vol.~14,
  no.~4, pp. 307--327, 2000.

\bibitem{laine2017temporal}
S.~Laine and T.~Aila, ``Temporal ensembling for semi-supervised learning,'' in
  \emph{ICLR}, 2017.

\bibitem{hong2021disentangling}
Y.~Hong, S.~Han, K.~Choi, S.~Seo, B.~Kim, and B.~Chang, ``Disentangling label
  distribution for long-tailed visual recognition,'' in \emph{CVPR}, 2021, pp.
  6626--6636.

\bibitem{hendrycks2022scaling}
D.~Hendrycks, S.~Basart, M.~Mazeika, A.~Zou, J.~Kwon, M.~Mostajabi,
  J.~Steinhardt, and D.~Song, ``Scaling out-of-distribution detection for
  real-world settings,'' in \emph{ICML}.\hskip 1em plus 0.5em minus 0.4em\relax
  PMLR, 2022, pp. 8759--8773.

\bibitem{klumpp2022common}
P.~Klumpp, T.~Arias-Vergara, P.~A. P{\'e}rez-Toro, E.~N{\"o}th, and J.~R.
  Orozco-Arroyave, ``Common phone: A multilingual dataset for robust acoustic
  modelling,'' \emph{arXiv preprint arXiv:2201.05912}, 2022.

\bibitem{zhao2018l2}
G.~Zhao, S.~Sonsaat, A.~Silpachai, I.~Lucic, E.~Chukharev-Hudilainen, J.~Levis,
  and R.~Gutierrez-Osuna, ``L2-arctic: A non-native english speech corpus.'' in
  \emph{Interspeech}, 2018.

\bibitem{kim2008uaspeech}
H.~Kim, M.~Hasegawa-Johnson, A.~Perlman, J.~Gunderson, T.~S. Huang, K.~Watkin,
  and S.~Frame, ``Dysarthric speech database for universal access research,''
  in \emph{Ninth Annual Conference of the International Speech Communication
  Association}, 2008.

\bibitem{choi2012dysarthric}
D.-L. Choi, B.-W. Kim, Y.-W. Kim, Y.-J. Lee, Y.~Um, and M.~Chung, ``Dysarthric
  speech database for development of qolt software technology.'' in
  \emph{LREC}, 2012, pp. 3378--3381.

\bibitem{ta2016dysarthric}
M.~TA, T.~Nagarajan, and P.~Vijayalakshmi, ``Dysarthric speech corpus in tamil
  for rehabilitation research,'' in \emph{Region TENCON}.\hskip 1em plus 0.5em
  minus 0.4em\relax IEEE, 2016, pp. 2610--2613.

\bibitem{enderby1980frenchay}
P.~Enderby, ``Frenchay dysarthria assessment,'' \emph{British Journal of
  Disorders of Communication}, 1980.

\bibitem{mcauliffe2017montreal}
M.~McAuliffe, M.~Socolof, S.~Mihuc, M.~Wagner, and M.~Sonderegger, ``Montreal
  forced aligner: Trainable text-speech alignment using kaldi.'' in
  \emph{Interspeech}, 2017, pp. 498--502.

\bibitem{babu2022xls}
A.~Babu, C.~Wang, A.~Tjandra, K.~Lakhotia, Q.~Xu, N.~Goyal, K.~Singh, P.~von
  Platen, Y.~Saraf, J.~Pino, A.~Baevski, A.~Conneau, and M.~Auli, ``{XLS-R:}
  self-supervised cross-lingual speech representation learning at scale,'' in
  \emph{Interspeech}, 2022.

\bibitem{choi2022opening}
K.~Choi and E.~J. Yeo, ``Opening the black box of wav2vec feature encoder,''
  \emph{arXiv preprint arXiv:2210.15386}, 2022.

\bibitem{loshchilov2019decoupled}
I.~Loshchilov and F.~Hutter, ``Decoupled weight decay regularization,'' in
  \emph{ICLR}, 2019.

\bibitem{hmm-gop}
J.~Zhang, ``Gmm-based gop (goodness of pronunciation) using kaldi.''
  \url{https://github.com/jimbozhang/kaldi-gop}, 2020.

\bibitem{mathad2021impact}
V.~C. Mathad, T.~J. Mahr, N.~Scherer, K.~Chapman, K.~C. Hustad, J.~Liss, and
  V.~Berisha, ``The impact of forced-alignment errors on automatic
  pronunciation evaluation.'' in \emph{Interspeech}, 2021, pp. 1922--1926.

\bibitem{hernandez2019acoustic}
A.~Hernandez, H.-y. Lee, and M.~Chung, ``Acoustic analysis of fricatives in
  dysarthric speakers with cerebral palsy,'' \emph{Phonetics and Speech
  Sciences}, vol.~11, no.~3, pp. 23--29, 2019.

\end{thebibliography}

\end{document}